\documentclass[12pt]{article}

\usepackage[dvipsnames]{xcolor}
\usepackage{amsmath}
\usepackage{amssymb}
\usepackage{dsfont}
\usepackage[english]{babel}
\usepackage[utf8]{inputenc}
\usepackage{times}
\usepackage{graphics}
\usepackage{graphicx}
\usepackage[T1]{fontenc}
\usepackage[official,right]{eurosym}
\usepackage{url}
\usepackage{eso-pic}

\usepackage[utf8]{inputenc}
\usepackage[title]{appendix}
\usepackage{algorithmicx}
\usepackage{algorithm}
\usepackage{algpseudocode}
\usepackage{amsmath,amssymb,amsfonts}
\usepackage{amsthm}
\usepackage{booktabs}
\usepackage{comment}
\usepackage{graphicx}
\usepackage{leftidx}
\usepackage{listings}
\usepackage{manyfoot}
\usepackage{mathrsfs}
\usepackage{multirow}
\usepackage{natbib}
\usepackage{pgfplots}
\usepackage{tensor}
\usepackage{textcomp}
\usepackage{tikz}
\usepackage{xcolor}
\usepackage{tabularx}
\usetikzlibrary{patterns}
\usetikzlibrary{plotmarks}
\usetikzlibrary{shapes.geometric}
\newcommand{\A}{\mathbf{A}}
\newcommand{\B}{\mathbf{B}}
\newcommand{\C}{\mathbf{C}}

\definecolor{green}{HTML}{34A853}
\definecolor{blue}{HTML}{4285F4}
\definecolor{yellow}{HTML}{FBBC05}
\definecolor{red}{HTML}{EA4335}

\newtheorem{prop}{Proposition}
\newtheorem{ax}{Property}
\newtheorem{ex}{Example}

\pgfplotsset{compat=1.18}

\title{A framework for expected capability sets}
\author{Nicolas Fayard$^1$, David Ríos Insua$^2$, Alexis Tsoukiàs$^1$ \\ $^1$CNRS-LAMSADE, PSL, Université Paris Dauphine\\
$^2$Institute of Mathematical Sciences (ICMAT)}
\date{}

\begin{document}

\thispagestyle{empty}

\enlargethispage*{8cm}
 \vspace*{-38mm}

\AddToShipoutPictureBG*{\includegraphics[width=\paperwidth,height=\paperheight]{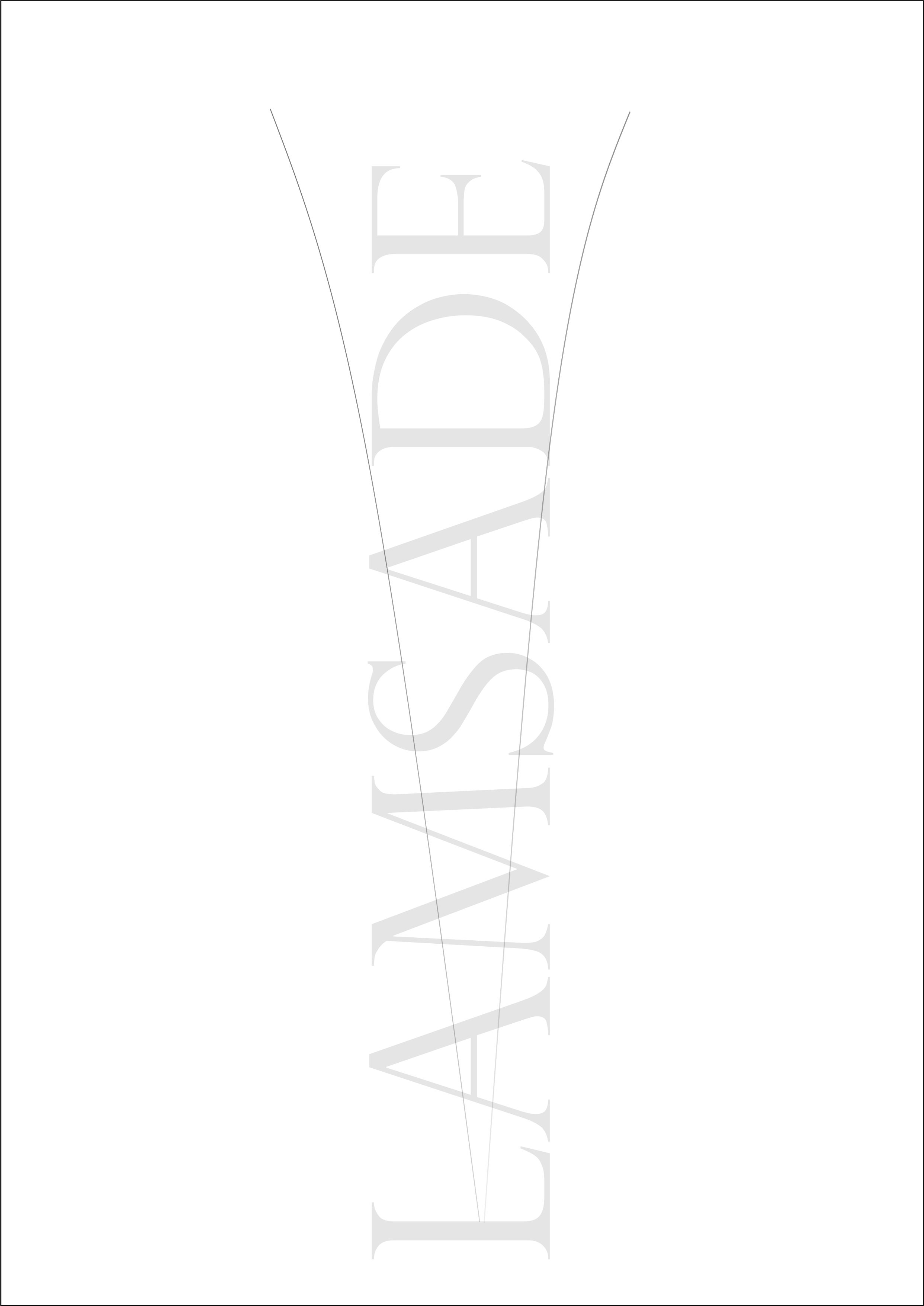}}

\begin{minipage}{24cm}
 \hspace*{-28mm}
\begin{picture}(500,700)\thicklines
 \put(60,670){\makebox(0,0){\scalebox{0.7}{\includegraphics{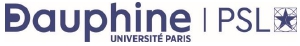}}}}
 \put(60,70){\makebox(0,0){\scalebox{0.3}{\includegraphics{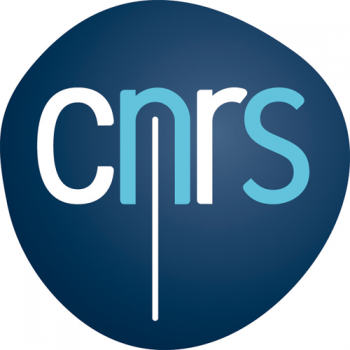}}}}
 \put(320,350){\makebox(0,0){\Huge{CAHIER DU \textcolor{BurntOrange}{LAMSADE}}}}
 \put(140,10){\textcolor{BurntOrange}{\line(0,1){680}}}
 \put(190,330){\line(1,0){263}}
 \put(320,310){\makebox(0,0){\Huge{\emph{407}}}}
 \put(320,290){\makebox(0,0){May 2024}}
 \put(320,210){\makebox(0,0){\Large{A framework for expected capability sets}}}
 \put(320,190){\makebox(0,0){\Large{}}}
 \put(320,100){\makebox(0,0){\Large{Nicolas Fayard, David Ríos Insua, Alexis Tsoukiàs}}}
 \put(320,670){\makebox(0,0){\Large{\emph{Laboratoire d'Analyse et Mod\'elisation}}}}
 \put(320,650){\makebox(0,0){\Large{\emph{de Syst\`emes pour l'Aide \`a la D\'ecision}}}}
 \put(320,630){\makebox(0,0){\Large{\emph{UMR 7243}}}}
\end{picture}
\end{minipage}

\newpage

\addtocounter{page}{-1}

\maketitle

\abstract{This paper addresses decision-aiding problems that involve multiple objectives and uncertain states of the world.
Inspired by the capability approach, we focus on cases where a policy maker chooses an act that, combined with a state of the world, leads to a set of choices for citizens.
While no preferential information is available to construct importance parameters for the criteria, we can obtain likelihoods for the different states.
To effectively support decision-aiding in this context, we propose two procedures that merge the potential set of choices for each state of the world taking into account their respective likelihoods.
Our procedures satisfy several fundamental and desirable properties that characterize the outcomes.}

\newpage

\section{Introduction.}\label{intro}

Consider the case of a policy maker deciding how to
allocate resources for climate change mitigation projects for a given
territory.

A simplified version of a decision analytic approach consists of assessing the current welfare of a territory based on the distribution of private assets and the accessibility of the common (resources). This approach involves identifying scenarios of potential damage to property and common resources, along with their likelihood of occurrence, and then solving a expected utility optimisation problem. From a decision support perspective, this implies that the welfare, consequences, and impacts of policies need to be measured on a single utility function that summarizes all these aspects.
As suggested in \cite{fayard2022capability, sen1985commodities, sen1993capability}, while this approach aids in rationalizing the decision-making process of policy makers, it fails to address, at least, two specific issues: \\
  - the impact of common resources upon the welfare of the
citizens; \\
  - the fact that the citizens do not behave as
indistinguishable consumers which means that considering the welfare as
just an utility maximisation problem is misleading. \\
In the aforementioned paper \citep{fayard2022capability}, the authors introduce the idea of using Sen's
``Capability Theory'' \citep{sen1979equality,sen1985commodities,sen1993capability,sen1999development,sen2009idea} to overcome both issues. Their approach consists of
assuming that the welfare of a single citizen is the Pareto frontier
resulting from solving a mixed integer multi-objective linear
programme (MI-MOLP) where constraints are given by the private assets of a
citizen as well as the accessibility to the common resources and, in turn, the objectives
are the maximisation of several value welfare functions. These functions reflect the individual's subjective preferences across different welfare dimensions, such as being safe, being well-nourished, being in good health, or being happy. 
The rationale for adopting this perspective is grounded on the significance of ``freedom of choice'' for each citizen. Essentially, welfare is determined not only by our possessions but also by the opportunities and capabilities that these possessions enable us to pursue.
When all significant welfare dimensions are taken into account, the
capability set can be perceived as a Pareto frontier, where, with no loss of generality,
we may assume that all such dimensions are to be maximized.

Once we know each citizen's capability set we can cluster the
population of
a given community according to their capability similarity (how similar are two Pareto
frontiers).
The consequence of this approach is that determining the welfare of a
community
(or a territory) is not just the sum of the utilities of the citizens
within that community.
Instead, it involves a Pareto frontier that represents multidimensional welfare for each cluster of citizens who share similar values.
As a result, the decision problem of the PM consists of establishing
(choosing) a
policy which maximises the capability of a (cluster of) citizen, which
means
maximising both the level (utility) of the outcomes (for each citizen)
and their
freedom of choice.

It is crucial to emphasize how the notion of capabilities refers to freedom. As an example from  \cite{sen1985commodities}, if a citizen has access to a capability set containing multiple solutions, among which option $b$ is considered to be the best, according to her preferences, a second capability set comprising only option $b$ would be deemed less desirable than the first set, even if the citizen would end up choosing option $b$ in both cases. This is because the second set provides less freedom, in the sense of facilitating fewer choices. When designing public policies, the question arises as to how to extend capability sets regarded as Pareto frontiers.
Any variation to the private assets of a citizen or their access to
the commons (such as damages to private property or to commons) will
result in modifying the Pareto frontier and, therefore, the citizen's
welfare. What we aim to model is how external events can affect the current welfare distribution (the capability sets) for different
clusters of citizens and how policies can mitigate such impacts.

Considering various events under different scenarios, we need to take
into account the varying likelihoods of such events occurring.
Assuming such likelihoods are measurable through probabilities, the
usual approach consists of computing the expected utilities. However,
for our policy maker aiming at maximising welfare for the citizens of her
territory there is no single utility function to consider for each
citizen, but a Pareto frontier representing the welfare supposed to be
achieved through the policies. The traditional expected public policy
approach results now in considering public policies as an Expected Pareto
frontier, where the Pareto frontiers resulting from different events
occurring are merged taking into account the likelihood of such events.
The open technical question is how to construct such ``Expected Pareto
frontiers''? To our knowledge this is an open question despite the
importance it has for policy design purposes.
\\

Thus, our normative position stands around the following policy design question: ``How can we design a policy that maximizes freedom of choice for citizens, focusing on actions that will invariably improve aspects of their lives?''. When multiple states of the world are plausible, and their likelihoods are known (even if they are subjective estimates), we aim to develop policies that maximize freedom of choice across all potential states of the world, factoring in their relative chance of occurrence.

This Capability Theory based perspectives differs from more classical approaches. 
For example, traditional decision-making methods like multi-attribute utility theory \citep{fishburn1977multiattribute, keeney1993decisions} aim to identify the \emph{best solution} for a decision-maker by quantifying
and aggregating multiple attributes of each alternative based on the maximum expected utility principle. 
Similarly, approaches at the juncture of scenario analysis and multi-criteria
decision making \citep{INSUA, durbach2012modeling} focus on the optimization of outcomes, but they do so across a collection of decisions rather than individual ones. They aim to balance the overall risk and return across a 'portfolio' of choices, taking into account potential correlations among them for optimal diversification and risk management. More specifically, we assume we have available probabilities for the states (differing from the traditional assumption
in scenario planning \citep{stewart2013integrating}) and  have capability sets 
rather than single consequences,  even if multi-attribute (differing from the standard 
scenario-based portfolio approaches \citep{INSUA, liesio2012scenario, vilkkumaa2018scenario}).
Our final discussion points out additional issues concerning partial information 
about the states' probabilities.

To summarize, the policy maker allocates resources within society, after which nature determines their actual distribution. We define the ways in which citizens can use these resources to experience specific types of lives as \emph{capability sets}.
Citizens then decide to choose a \emph{beings} (seen as states of existence encompassing various dimensions of human welfare) from their capability set based on their individual values.
The manner in which citizens use the resources and the beings they choose are personal choices.
The objective of the policy maker is to ensure that citizens have access to a diverse and appreciated capability set.
The policy maker must consider the resources they allocate, as well as those determined by natural forces.
Should probabilities of nature’s actions be available,
the policy maker might then adopt a policy that optimizes the aggregated capability
set, as represented by the Pareto Frontier.

In this paper, our primary focus is on the exploration of mixed capability sets within the previously outlined framework. We define a mixed capability set as the combination or aggregation of capability sets corresponding to different states of the world, each weighted by their respective subjective probabilities.
To achieve this, we begin by precisely formulating the problem in Section~\ref{terminology}, where we introduce the necessary terminology.  Section~\ref{model_mix} introduces a natural approach toward mixing capability sets, termed \emph{average capability sets}. In Section~\ref{sec:exp}, we propose a more sophisticated approach called the \emph{expected capability set}. Section~\ref{properties} delves into the major properties of the expected capability set. Section~\ref{seq:vs} further explores the distinctions between both approaches, expanding our understanding of the possibilities and considerations involved in decision-making within the capability approach context.

\section{Problem Formulation.}\label{terminology}

A Policy Maker (PM) is faced with the task of selecting an act in an environment of risk. To represent this risk, a set $S$ of states of the world is employed.
$S$ is assumed to be finite, with $S=\{s_1, s_2,..., s_{l^*}\}$. The PM's beliefs about the likelihood of each state being the actual one can be modeled using subjective probabilities $p(s_l)$ for $l\in \{1, 2, \cdots , l^*\}$, in accordance with well-established results from the literature such as \cite{degroot,french,scott1964measurement}, with $\sum_{l=1}^{l^*}p(s_l)=1$ and $p(s_l)\geq0$ for all $s_l$.

The PM has to select an act from a set $F=\{f_1, f_2, ..., f_{m^*}\}$ which maps states from $S$ into consequences. To evaluate these consequences, we employ a function $U$, resulting in a capability set \(U(f_m(s_l)) \subseteq \mathbb{R}^{h^*}_+\) for each \(m \in \{1, 2, \ldots, m^*\}\) and \(l \in \{1, 2, \ldots, l^*\}\). 
The \emph{beings} within these sets, denoted as \(\vec{b} \in \{U(f_m(s_l)) \,|\, \forall (f_m, s_l) \in F \times S\}\), are \(h^*\)-dimensional vectors that encapsulate various welfare dimensions such as health, security, pleasure, and more.
It is essential to acknowledge that capability sets are generally not singletons and are guaranteed to be non-empty (\(U(f_m(s_l)) \neq \emptyset\)), potentially encompassing an infinite number of beings,  and must form a compact (closed and bounded) set.
For the sake of simplicity in notation, we will represent sure acts \(U(f_m(S)), U(f_{m^\prime}(S)), U(f_{m^{\prime\prime}}(S)), \ldots\) as \(\A, \B, \C, \ldots\) respectively. This notation will be utilized to discuss  capability sets in general terms. Similarly, \(U(f_m(s_l)), U(f_{m^\prime}(s_l)), U(f_{m^{\prime\prime}}(s_l))\) will be denoted as \(\A_l, \B_l, \C_l\).


It is important to note that the choice of \emph{beings} $\vec{b}$ within $\A_l$, made by the citizen after the PM selects $f_m$ and nature determines $s_l$, is not the concern to the PM. The PM intentionally avoids making assumptions about which beings will actually be chosen for two primary reasons: first, the PM may not have access to the individual's preferences across the various dimensions of welfare; second, the objective of the PM is to enhance freedom without dictating a specific lifestyle.
Indeed, the \emph{sole} assumption the PM makes about the citizen is that all welfare dimensions relevant to the citizen's choice are included in the $h^{*}$ welfare dimensions, and that assuming increasing monotonic preferences over welfare, the citizen is rational. Thus, we will choose a $\vec{b}$ in \(\A_l\) that is also in \(PF(\A_l)\), where $PF$ represents the Pareto frontier, as in \cite{fayard2022capability}.
When a capability set \(\A_l\) comprises a finite number $n_l^*$ of solutions, we represent its beings by \(\vec{b_{l, 1}}, \ldots ,\vec{b_{l, n_l^*}}\)  (where \(b_{l, n, h}\) specifies the value of the \(n^{th}\) beings for the \(l^{th}\) state in the \(h^{th}\) dimension).

We define \(\A - \mathbb{R}^{h^*}_+\) as the set encompassing all beings weakly Pareto dominated by \(\A\); that is, it includes all beings \(\vec{b}\) for which there exists a \(\vec{b}^\prime \in \A\) such that \(b_h \leq b_h^\prime\) for all dimension $h$, denoted \(\vec{b} \leq \vec{b}^\prime\).\footnote{Note that \(\vec{b} = \vec{b}^\prime \implies \vec{b} \leq \vec{b}^\prime \text{ and } \vec{b}^\prime \leq \vec{b}\)}

One method to determine whether a capability set \(\B\) is preferred to a capability set \(\A\), when the state of nature is known, is to verify that for each beings \(\vec{b} \in \A\), there is a beings \(\vec{b^\prime} \in \B\) such that \(\vec{b} \leq \vec{b^\prime}\).
Additionally, there should be at least one $\vec{b^\prime} \in \B$ for which there is no $\vec{b} \in \A$ that is at least as good as $\vec{b^\prime}$. 
 In other words, if $\B$ is ``above'' $\A$, or $\A\subseteq \B-\mathbb{R}^{h^*}_{+, *}$, then $\B$ is preferred to $\A$, and if $\A\subseteq \B-\mathbb{R}^{h^*}_+$ then $\B$ is at least as good as $\A$. 
 This ensures that any rational individual, assuming they aim to maximize their well-being across all dimensions, would choose $\B$ over $\A$, regardless of their specific preferences.\\

\begin{ex}
Figure \ref{fig:Three_Capability_set} depicts three compact capability sets with $h^* = 2$. Two of the capability sets, respectively denoted $\A$ and $\B$, are finite, while the third 
one, denoted $\C$, contains an infinite number of solutions. The areas in the plot represent the spaces dominated by $\A$ and $\B$, specifically $\A - \mathbb{R}^2_+$ and $\B - \mathbb{R}^2_+$ (limited to the first quadrant).

\begin{figure}[hbt]
        \centering
        \begin{tikzpicture}
\begin{axis}[
    title={},
    xlabel={Criterion $1$},
    ylabel={Criterion $2$},
    xmin=0, xmax=13.5,
    ymin=0, ymax=10.5,
    grid=minor,
    axis x line=bottom, axis y line = left,
    legend style={at={(1.03,1)},anchor=north},
    legend cell align={left}]
  \draw [gray!20,fill=gray!20] (0,0) --(12, 0)--(12, 1)--(10, 1)--(10, 2)--(5, 2)--(5, 3)--(4, 3)--(4, 6)--(3.5, 6)--(3.5, 7.5)--(2, 7.5)--(2, 8)--(0, 8) --(0,0)-- cycle;
 \draw[pattern=dots] (0,0) --(13, 0)--(13, 2)--(11, 2)--(11, 3)--(5, 3)--(5, 9)--(0, 9) --(0,0)-- cycle;
 	
\addplot[fill=black!5, mark = *, only marks, mark size = 2.5pt]
coordinates {(12, 1)(10, 2)(5, 3)(4, 6)(3.5, 7.5)(2, 8)};

\addplot[fill=black!40, mark = *, only marks, mark size = 2.5pt]
coordinates {(13, 2)(11, 3)(5, 9)};
\addplot[line width=1pt]
coordinates {(9, 0)(8, 7.5)(0, 8.5)};
\addlegendimage{area legend,fill=gray!20}
\addlegendimage{area legend,pattern=dots}
\addplot[dashed]
coordinates {(12, 0)(12, 1)(10, 1)(10, 2)(5, 2)(5, 3)(4, 3)(4, 6)(3.5, 6)(3.5, 7.5)(2, 7.5)(2, 8)(0, 8)};
\addplot[dotted]
coordinates {(13, 0)(13, 2)(11, 2)(11, 3)(5, 3)(5, 9)(0, 9)};

\addlegendentry{$\A$};
\addlegendentry{$\B$};
\addlegendentry{$\C$};


\addlegendentry{$\A - \mathbb{R}^2_+$};
\addlegendentry{$\B - \mathbb{R}^2_+$};
\end{axis}
\end{tikzpicture}
        \caption{Three capability sets $\A, \B, \C$}
        \label{fig:Three_Capability_set}
    \end{figure}
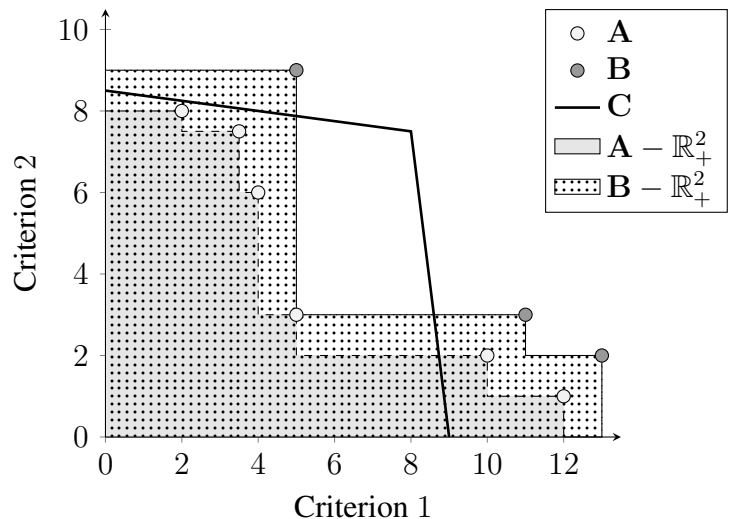
    
In this example, $\B$ should be preferred to $\A$ because $\A$ is entirely contained within $\B - \mathbb{R}^2_+$. On the other hand, when comparing $\C$ to $\A$ using the above method, no comparison can be made. This is because neither $\C$ is a subset of $\A - \mathbb{R}^2_+$, nor is $\A$ a subset of $\C - \mathbb{R}^2_+$.\hfill $\triangle$  
\end{ex}

\vspace{0.1in}

Under classical assumptions \citep{savage1972foundations}, decision support within $F$ is carried out as follows. For a given act and state of the world, there exists a unique consequence  $f_m(s_l)$. There is also a utility function $u$ such that $u(f_m(s_l)) \in \mathbb{R}$ for all $f_m\in F$ and $s_l\in S$, such that $f_{m^\prime}$ is at least as preferred as $f_{m}$ 
if and only if the expected utility of $f_{m^\prime}$ is not less than that of $f_{m}$, i.e., $$E(f_{m'}) = \sum_{l=1}^{l^*} p(s_l) \cdot u(f_{m'}(s_l)) \geq  \sum_{l=1}^{l^*}  p(s_l) \cdot u (f_{m}(s_l)) = E(f_{m}). $$ Consequently, the PM is recommended 
to select the act maximizing expected utility, namely $$\max_{f_m\in F}(E(u(f_m)).$$


Ultimately, the major difference in our framework is that we use capability sets \(U(f_m(s_l)) \subseteq \mathbb{R}^{h^*}_+\), instead of utilities (i.e., \(u(f_m(s_l)) \in \mathbb{R}\)).
Subsequent sections in this paper aim to elucidate the manner in which an expected capability set can be established.

\section{Average and Expected capability sets}
In this section, we propose two ways to mix capability sets. One approach, referred to as the \emph{average capability set} seems quite natural, yet does not respect some desirable properties. The other one, known as the \emph{expected capability set}, is more sophisticated and generally more relevant to our problem at hand.

\subsection{Average capability set}
\label{model_mix}
A natural approach to addressing this problem,  involves considering every combination of beings from various capability sets. These combinations are then aggregated based on the probability of their corresponding states of the world. This method of aggregation is referred to as the \emph{average capability set}, denoted by $\overline{\A}$ for an act $f_m$:

\begin{align*}
\overline{\A} = \Biggl\{  \sum_{l=1}^{l^*} p(s_l) \cdot X_{i, l}, 
 \text{for all } X_i \text{ such that } 
 X_i = (\vec{b_1}, \ldots , \vec{b_{l^*}}) \\
 \text{ with } \vec{b}_l \in \A_l 
 \text{ for every } l\in \{1, \ldots , l^*\} \Biggr\}.
\end{align*}
It is a natural extension of the expected utility concept and is consistent with it: if the capability sets
include just one beings and are only assessed using one dimension, then expected
capability sets are equivalent to expected utilities (see proof in the appendix). 
\noindent If our goal is to identify and retain only the efficient Pareto set of the set $\overline{\A}$, denoted as $PF(\overline{\A})$,  we calculate it by solving Problem \ref{eq:ag}

\begin{equation}\label{eq:ag}
\begin{array}{ll@{}ll@{}llll}
PF(\overline{\A}) = PF & \displaystyle &  \Bigl(\sum_{l=1}^{l^*} b_{l,1} \cdot p(s_l), \cdots, &\sum_{l=1}^{l^*} b_{l, h^*} \cdot p(s_l)\Bigr)\\
\text{s.t.}\\
\displaystyle & &\vec{b}_l \in \A_l&  \forall l \in \{1,\cdots, l^*\}.  \\
\end{array}
\end{equation}
Generally, reducing to the Pareto frontier is not mandatory. However, it serves as a step in elucidating the concept of the expected capability set and facilitates the comparison of both approaches. The decision to use or not the Pareto frontier will be discussed in Section \ref{seq:vs}.

\vspace{.05in}

 \begin{ex}\label{ex:av}
 Consider a case where $U(f_m(s_1))= \A_1= \{(2,7), (3,4)\}$ and $U(f_m(s_2))=\A_2 =\{(4,3), (7,2)\}$. Should $s_1$ be the actual state, a citizen would choose between solutions $(2,7)$ and $(3,4)$, whereas if $s_2$ was the actual one, such citizen would choose between $(4,3)$ and $(7,2)$.
 With $p(s_1)=p(s_2)=0.5$, the average capability set $\overline{\A}$ is 
$\{(3,5), (3.5,3.5), (4.5,4.5), $ $(5,3)\}$ and its Pareto frontier $PF(\overline{\A})$ is 
$\{(3,5), $$(4.5,4.5), (5,3)\}$. Refer to Table \ref{tab:average} for 
the steps used to construct this set and Figure \ref{fig:E5} for a visual depiction.
\begin{table}[ht]
    \centering
    \begin{tabular}{l|l}
         $X_i$& $\vec{b} \in \overline{\A}$ \\
         $\{(2, 7),(4, 3)\}$&$\mathbf{(3, 5)}$\\
         $\{(2, 7),(7, 2)\}$&$\mathbf{(4.5, 4.5)}$\\
         $\{(3, 4),(4, 3)\}$&$(3.5, 3.5)$\\
         $\{(3, 4),(7, 2)\}$&$\mathbf{(5, 3)}$\\
    \end{tabular}
    \caption{Construction of solutions of $\overline{\A}$}
    \label{tab:average}
\end{table}

\noindent It is important to note that this mixed solution includes the possibility of achieving $(3.5, 3.5)$, which is dominated by $(4.5, 4.5)$. Additionally, in this particular case, the average capability set is not dominated by any beings from $\A_1$ or $\A_2$. In fact, no beings within the average capability set is dominated by beings from either $\A_1$ or $\A_2$.\hfill $\triangle$ 
\vspace{.05in}
\vspace{.05in}
\begin{figure}
\centering
\begin{tikzpicture}
\begin{axis}[
    title={},
    xlabel={Criterion 1},
    ylabel={Criterion 2},
    xmin=0, xmax=10,
    ymin=0, ymax=10 ,
    xtick={0, 1, 2, 3, 4, 5, 6, 7, 8, 9},
    ytick={0, 1, 2, 3, 4, 5, 6, 7, 8, 9},
    grid=major,
    axis x line=bottom, axis y line = left,
    legend style={at={(1.03,1)},anchor=north},
    legend cell align={left}]

\addplot[fill=black!90, mark = triangle*, only marks] coordinates {(2, 7)(3, 4)};
\addplot[fill=black!90, mark = square*, only marks] coordinates {(7,2)(4,3)};
\addplot[fill=black!50, mark = *, only marks] coordinates {(4.5,4.5)(3,5)(5,3)(3.5,3.5)};
\addlegendentry{$\A_1$};
\addlegendentry{$\A_2$};
\addlegendentry{$\overline{\A}$};
\draw (20,70)--(70,20);
 \draw (20,70)--(40,30);
 \draw (30,40)--(70,20);
 \draw (30,40)--(40,30);

\end{axis}

\end{tikzpicture}
    \caption{Example \ref{ex:1}: Average capability set}
    \label{fig:E5}
\end{figure}
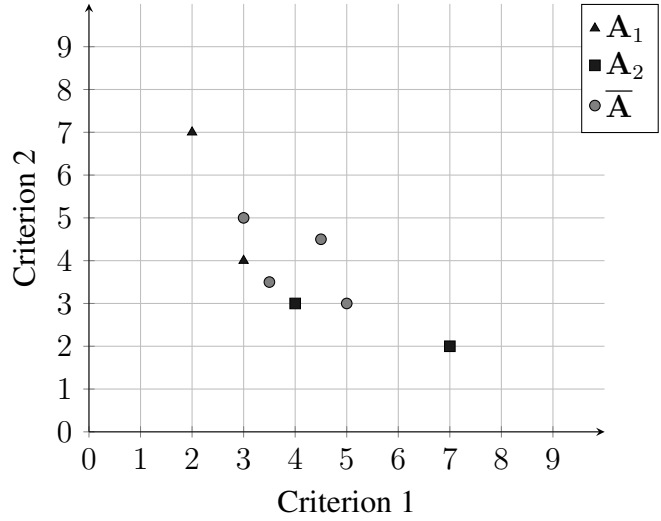

\end{ex}

Although the average capability set provides a straightforward method for aggregating sets of capabilities, it will prove inadequate in our problem. In the following subsection, we demonstrate the need to impose stronger properties on such a procedure, which we will refer to as the \emph{expected capability set}. Nevertheless, as Section \ref{seq:vs} explores, the average capability set can be useful for a problem that slightly differs from ours.

\subsection{Expected capability set}\label{sec:exp}
Let us introduce two properties that naturally capture any citizens' preferences in
relation to capability sets that will inform our discussion.
The first one, called \emph{greater choice} property, reflects the fundamental property within the 
capability approach that citizens prefer to have access to more and better beings.
 \vspace{.05in}
 
\begin{ax}[Greater choice]\label{ax:union}
For all capability sets $\A$ and $\B$, it holds that $\A \cup \B$ is at least as preferred as $\A$.
\end{ax} 
 \vspace{.05in}
 
\noindent 
By iteratively applying this property, 
we find  that the union of the capability sets resulting from an act $f_m$ across all the states, denoted  $\cup_{l=1}^{l^*} \mathbf{A_l}$, should be not less preferred to the capability set corresponding to each individual state. 
Accordingly, we should also expect that such union will also be not less preferred to the expected capability set, to be denoted $E(\A)$.

The second one, designated the {\em fewer choice} property, captures another aspect of citizens' preferences: for any capability set, citizens do not prefer having less choice.
\vspace{.05in}

\begin{ax}[Fewer choice]\label{ax:inter}
For all capability sets $\A$ and $\B\subseteq \mathbb{R}^{h^*}_+$, it holds that $\A$ is at least as preferred as $(\A - \mathbb{R}^{h^*}_+)\cap (\B -\mathbb{R}^{h^*}_+)$.
\end{ax}
\vspace{.05in}

\noindent Based on this property, the capability set associated with each individual state 
should not be less preferred to the intersection of the solutions
dominated by the capability sets resulting from an act $f_m$ across all states, that is $\cap_{l=1}^{l^*} (\A_l - \mathbb{R}^{h^*}_+)$. Accordingly, the expected capability sets $E(\A)$ should be not less preferred than such set. 

A simple example illustrates these properties.
\vspace{.05in}
\addtocounter{ex}{-1}
\begin{ex}[Cont]\label{ex:1}
Figure \ref{fig:E1} depicts the capability sets $\A_1$ and $\A_2$, as well as the space dominated by the solutions in the union of the capability sets under both 
states, denoted $(\A_1- \mathbb{R}^2_+)\cup (\A_2 - \mathbb{R}^2_+)$, in light gray, and the space that can be dominated in any state, denoted by $(\A_1 - \mathbb{R}^2_+) \cap (\A_2 - \mathbb{R}^2_+)$, in dark gray.
The average capability set is not contained within the light gray area and does not fulfill Property \ref{ax:union}.

Following Property \ref{ax:union}, the expected capability set $E(\A)$ should be contained within the light gray area, which is not the case for the average capability set. The space above the light gray area represents the beings that we are sure not to be able to dominate, regardless of the actual state of the world. On the other hand, following Property \ref{ax:inter}, the dark gray area should be dominated by $E(\A)$, 
 since in any state, we should find a solution dominating such area and,
 therefore, are sure to be able to dominate it.
\hfill $\triangle$  
\vspace{.05in}
\vspace{.05in}

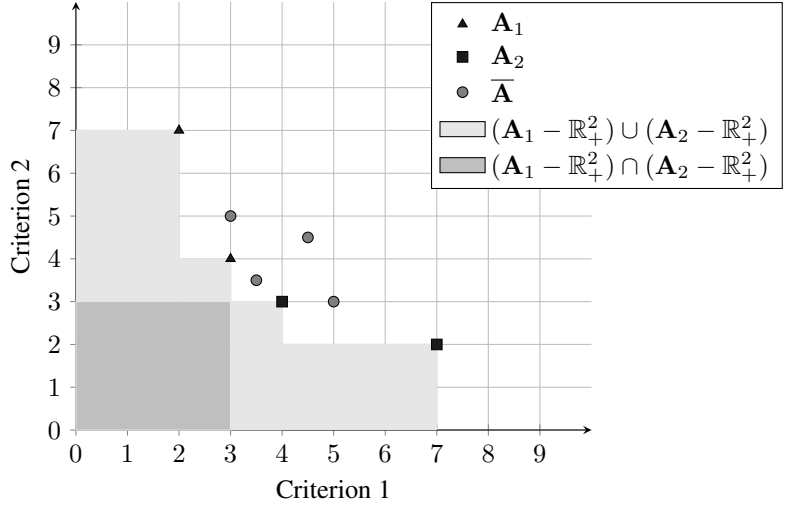
\begin{figure}[htbp]
\centering
\begin{tikzpicture}
\begin{axis}[
    title={},
    xlabel={Criterion 1},
    ylabel={Criterion 2},
    xmin=0, xmax=10,
    ymin=0, ymax=10 ,
    xtick={0, 1, 2, 3, 4, 5, 6, 7, 8, 9},
    ytick={0, 1, 2, 3, 4, 5, 6, 7, 8, 9},
    grid=major,
    axis x line=bottom, axis y line = left,
    legend style={at={(1.03,1)},anchor=north},
    legend cell align={left}]
 \draw  [gray!20,fill=gray!20] (0.01,0.01) -- (0.01,2) -- (7,2)--(7,0.01) -- cycle;
  \draw [gray!20,fill=gray!20] (0.01,0.01) -- (0.01,3) -- (4,3)--(4,0.01) -- cycle;
\draw [gray!20,fill=gray!20] (0.01,0.01) -- (0.01,3) -- (4,3)--(4,0.01) -- cycle;
 \draw [gray!20,fill=gray!20] (0.01,0.01) -- (0.01,7) -- (2,7)--(2,0.01) -- cycle;
  \draw [gray!20,fill=gray!20] (0.01,0.01) -- (0.01,4) -- (3,4)--(3,0.01) -- cycle;
    \draw[gray!20, fill=gray!50] (0,0) rectangle (3,3);
\draw[gray!50] (2,0)--(2,3);
\draw[gray!50] (0, 2)--(3,2);
\addplot[fill=black!90, mark = triangle*, only marks] coordinates {(2, 7)(3, 4)};
\addplot[fill=black!90, mark = square*, only marks] coordinates {(7,2)(4,3)};

\draw (8.5,-4) node {$u_1$};
\draw (-4,8.5) node {$u_2$};
\addplot[fill=black!50, mark = *, only marks] coordinates {(4.5,4.5)(3,5)(5,3)(3.5,3.5)};
\footnotesize
\addlegendentry{$\A_1$};
\addlegendentry{$\A_2$};
\addlegendimage{area legend,fill=gray!20}
\addlegendimage{area legend,fill=gray!50}
\addlegendentry{$\overline{\A}$};
\addlegendentry{$(\A_1 - \mathbb{R}^2_+) \cup (\A_2 - \mathbb{R}^2_+)$};
\addlegendentry{$(\A_1 - \mathbb{R}^2_+) \cap (\A_2 - \mathbb{R}^2_+)$};

\end{axis}
\end{tikzpicture}
    \caption{Example \ref{ex:1} used to illustrate Axioms 1 and 2.}
    \label{fig:E1}
\end{figure}

\end{ex}
Our objective is thus to establish a mixing procedure that ensures the condition 
\[
\bigcap_{l=1}^{l^*} (\A_l - \mathbb{R}^{h^*}_+) \subseteq E(\A) - \mathbb{R}^{h^*}_+ \subseteq \bigcup_{l=1}^{l^*} \A_l - \mathbb{R}^{h^*}_+
.\]
Note that this condition is not a definition of the expected capability set \( E(\A) \), but rather a crucial requirement that such set must fulfill to be considered effective in our context.
To operationalize it, we propose a model ensuring that for any \( \vec{b} \in E(\A) \), there exists \( \vec{b}^\prime \in \bigcup_{l=1}^{l^*} \A_l \) such that \( \vec{b} \leq \vec{b}^\prime \) (i.e., \( b_h \leq b_h^\prime \) for all \( h \)). Additionally, for any \( \vec{b}^{\prime\prime} \) in \( \bigcap_{l=1}^{l^*} \A_l - \mathbb{R}^{h^*}_+ \), there exists \( \vec{b} \in E(\A) \) such that \( \vec{b}^{\prime\prime} \leq \vec{b} \).

To do this, we shall provide several mathematical programming formulations to define and compute $E(\A)$. The first formulation, referred to as Problem \ref{eq:model}, is designed to define the expected capability set. Within this framework, beings $\vec{b}_l$ are not contained in $\mathbf{A}_l$ as in the averaging Problem \ref{eq:ag}, but are just weakly Pareto dominated by $\mathbf{A}_l$ and are combined, using the probabilities associated with their specific states as in Problem \ref{eq:ag}. Additionally, a necessary condition for aggregation is imposed: to preserve weak Pareto dominance across beings of in different states. This process ensures that the resulting expected capability set fulfills the desired two properties.
Specifically, the problem is formulated as
\begin{subequations}
\label{eq:model}
\begin{align} 
    E(\A)&=PF\Biggl(\sum_{ l=1}^{l^*} b_{l, 1} \cdot p(s_l), \ldots& \ldots, \sum_{ l=1}^{l^*}  b_{l, h^*}\cdot p(s_l)\Biggr)\nonumber\\
&
\vec{b}_l\in \A_l- \mathbb{R}^{h^*}_+ & \forall l\in \{1,\cdots, l^*\}\label{equ:prem}\\
&\vec{b_l} \geq \vec{b_{l^\prime}} \, \text{ or } \, \vec{b_l} \leq \vec{b_{l^\prime}} & \forall l, l^\prime \in \{1, \ldots, l^* \}, \;\; l\neq l^\prime.\label{equ:deux}
\end{align}
\end{subequations}
Condition (\ref{equ:prem})  directs our focus towards mixing all beings \emph{dominated} by $\A_l$ for all $s_l\in S$, rather than just the beings of $\A_l$. On the other hand, 
 condition (\ref{equ:deux})  stipulates the existence of a total order among all $\vec{b}_l$ 
 (whereby $\vec{b}_l\geq \vec{b}_{l^\prime}$ if and only if $b_{l,h}\geq b_{l^\prime,h}$ for all $h\in \{1,\cdots, h^*\}$). 
This implies that any two beings $\vec{b_l}\in \A_l, \vec{b_{l^\prime}}\in \A_{l^\prime}$ 
 which are incomparable (i.e., neither $\vec{b}_l\geq \vec{b}_{l^\prime}$ nor $\vec{b}_l\leq \vec{b}_{l^\prime}$) are inappropriate to lead to an expected solution of the type $\vec{b}=p(s_l)\cdot \vec{b}_{l}+ p(s_{l^\prime}) \cdot \vec{b}_{l^\prime}$.  Note that in the classical settings \citep{savage1972foundations} with one-dimensional utilities ($h^*=1$) and a single beings in the capability sets, all beings are comparable as $\geq$ is a total order on  $\mathbb{R}$. 



\noindent


While Problem \ref{eq:model} has been employed to elucidate the concept of expected capability set, we still need to provide computational schemes to build them. When dealing with a collection of compact capability sets \( \A_l \) (with the different \( \A_l \) possibly obtained through a MI-MOLP as in \cite{fayard2022capability}), we turn to a second, more operational formulation to derive \( E(\A) \). It is based on a MI-MOLP, referred to as Problem \ref{eq:realform}, where \( M \) represents a suitably large number.

\begin{subequations}
\label{eq:realform}
\begin{align}
&E(\A)=PF\Biggl(\sum_{ l=1}^{l^*} b_{l, 1} \cdot p(s_l), \ldots& \ldots, \sum_{ l=1}^{l^*}  b_{l, h^*}\cdot p(s_l)\Biggr)\nonumber\\
\text{s.t.}\nonumber\\
& b_{l, h} \leq z_{l, h} &\forall l \in \{1, \cdots , l^*\}\label{one}\\
&& \forall h \in \{1, \ldots , h^*\}\nonumber\\
& \vec{z_l} \in \A_l & \forall l \in \{1, \ldots, l^*\}\label{dom}\\
& d_{l, l^\prime} \in \{0, 1\}& l\neq l^\prime\;\; l, l^\prime \in \{1, \cdots , l^*\} \label{decvar}\\
& b_{l, h} \leq b_{l^\prime, h} + d_{l, l^\prime} \cdot M  & l\neq l^\prime\;\; l, l^\prime \in \{1, \cdots , l^*\} \label{biger}\\
& d_{l, l^\prime} + d_{l^\prime, l} \leq 1 & l< l^\prime\;\; l, l^\prime \in \{1, \cdots , l^*\} \label{allsol}\\
& b_{l, h} \in \mathbb{R}& \forall  l \in \{1, \cdots , l^*\}\\
&& \forall h \in \{1, \cdots , h^*\}\nonumber 
\end{align}
\end{subequations}


\noindent
 In this formulation, inequality (\ref{one}) guarantees that we aggregate vectors $\vec{b_l}$ that are dominated by a solution $\vec{z_l}$ in $\A_l$, as expressed by (\ref{dom})
 which compiles the constraints to find the capability set of $\mathbf{A}_l$, as in \cite{fayard2022capability}.
To establish a total order among the solutions $\vec{b}_l=(b_{l, 1}, \ldots, b_{l, h^*})$, we introduce binary decision variables $d_{l,l^\prime}$ using Equations (\ref{decvar}). These variables ensure that if $d_{l, l^\prime}=0$, then $b_{l, h} \leq b_{l^\prime, h}$ for all $h$, as in inequality (\ref{biger}). We ensure that this order exists between all solutions $\vec{b_l}$ by using inequality (\ref{allsol}).

Two key factors influence the computational complexity of Problem \ref{eq:realform}.
First, the \emph{number $l^*$ of states of the world} plays a significant role. 
Indeed, $\sum^{l^*-1}_{l=1}l\times 2$ decision variables $d_{l, l^\prime}$
are introduced to establish a total order between the weakly Pareto dominated solutions in each state. Thus, the bigger 
the number of states, the more decision variables and constraints need to be considered.
  Second, the \emph{number $h^*$ of welfare dimensions} affects the complexity of finding the expected Pareto frontier, a crucial aspect in our framework. When the dimensionality of the criteria is large, computational demands become significant, scaling typically  exponentially with the number of criteria \citep{ehrgott2005multicriteria}.

 If the capability sets $\A_l$ are finite $\{\vec{z_{l, 1}},\cdots , \vec{z_{l, n_l^*}}\}$, the expected capability set $E(\A)$ can be found through the following MI-MOLP (Problem \ref{eq:deuxieme}).
 Where, again, $M$ is a sufficiently large number 
\begin{subequations}
\label{eq:deuxieme}
\begin{align}
&E(\A)=  PF\Biggl(\sum_{\forall l=1}^{l^*} b_{l, 1} \cdot p(s_l),& \cdots, \sum_{\forall l=1}^{l^*} b_{l, h^*}\cdot p(s_l)\Biggr)\nonumber\\
\text{s.t.}\nonumber\\
& \sum^{n^*_l}_{n=1} \delta_{l,n} \leq n^*_l -1 & \forall l \in \{1, \cdots , l^*\}\label{sumdelta}\\
& b_{l, h} \leq z_{l,n,h} + \delta_{l,n} \cdot M & \forall l\in \{1,\cdots,l^*\}, \label{onebis} \\
&& \forall n \in \{1, \cdots , n_l^*\}\nonumber\\
&& \forall h \in \{1, \cdots , h^*\}\nonumber\\
& d_{l, l^\prime} \in \{0, 1\}& l\neq l^\prime\;\; l, l^\prime \in \{1, \cdots , l^*\}\label{decvarbis}\\
& b_{l, h} \leq b_{l^\prime, h} + d_{l, l^\prime} \cdot M 
& l\neq l^\prime\;\; l, l^\prime \in \{1, \cdots , l^*\} 
\label{bigerbis}\\
& d_{l, l^\prime} + d_{l^\prime, l} \leq 1 & l<l^\prime\;\; l, l^\prime \in \{1, \cdots , l^*\} \label{allsolbis}\\
& \delta_{l,n} \in \{0, 1\}& \forall l \in \{1, \cdots, l^*\} \label{delta}\\
&& \forall n \in \{1, \cdots, n_l^*\}\nonumber\\
& b_{l, h} \in \mathbb{R}& \forall  l \in \{1, \cdots , l^*\}\\
&& \forall h \in \{1, \cdots , h^*\}\nonumber\nonumber 
\end{align}
\end{subequations}
Constraints (\ref{decvarbis}), (\ref{bigerbis}) and (\ref{allsolbis})  establish an order among the aggregated beings as with constraints (\ref{decvar}), (\ref{biger}), and (\ref{allsol}). To determine dominated beings, we no longer use constraints (\ref{one}) and (\ref{dom}) but rather introduce new 
  constraints  (\ref{sumdelta}), (\ref{onebis}) and (\ref{delta}).
We introduce decision variables $\delta_{l,n}$ using constraint (\ref{delta}). Constraint (\ref{dom}) is now replaced by (\ref{onebis}). If $\delta_{l, n}$ equals $0$, then the beings $\vec{b_l}=(b_{l, 1}, \ldots , b_{l, h^*})$ is considered weakly dominated by the $n-{th}$ beings in the capability set $\A_l$. 
Finally, constraint (\ref{sumdelta}) ensures that at least one of the decision variables $\delta_{l,n}$ equals $0$, indicating that $\vec{b_l}$ is dominated by at least one solution in $\A_l$.

In addition to the number of states and the number of dimensions, a third factor influences the computational complexity of Problem \ref{eq:deuxieme}.
The \emph{size of the  capability sets} affects computational complexity. We introduce $\sum_{l=1}^{l^*} n_l^*$ decision variables $\delta_{l, n}$, where $n^*_l$ represents the number of solutions in each capability set $\A_l$. Thus, larger capability sets involve more decision variables and constraints, increasing computational demands.

Our proposal provides valuable insights into decision aiding under uncertainty in the context of the capability approach. However, it is important to consider carefully the above
computational limitations 
when applying the proposed framework. Addressing these limitations presents a compelling line of research that could significantly improve the tractability of our model.   As a last resort, we would need to draw on multiobjective heuristics as in  \cite{branke2016mcda,chugh2019survey}, or \cite{deb2002fast}.
\vspace{.05in}

\begin{ex} \label{para:ex2}
Consider an act $f_m$ with two states: $U(f_m(s_1)) = \A_1 = \{(3, 10), (4, 5), (7, 3), (8, 1)\}$ and $U(f_m(s_2))  = \A_2 =  \{(2, 5), (5, 4), (10, 2)\}$. 
    Figure \ref{fig:example_mix} displays $\A_1 $, $\A_2 $, and
    $\A_1- \mathbb{R}^2_+$ and $\A_2 - \mathbb{R}^2_+$ as 
    all the points on and under the dotted (resp. dashed) lines.\footnote{For ease of representation,
    the figure only displays the non-negative region.}
    It also shows $E(f_m)$ according to the formulation given in Problem \ref{eq:deuxieme},
    when $p(s_1)=0.8$ (and $p(s_2)=0.2$). 
We obtain $E(f_m)=\{(2.8, 9),$
$(3, 8.8),$
$(4, 4.8),$
$(4.2, 4),$
$(5, 3.2),$
$(6.6,3),$
$(7, 2.8),$
$(7.6,2),$
$(8.4, 1.2)\}$. 
As an example, $(2.8, 9)$ is obtained by aggregating $\vec{b}_1 = (3, 10)$  and $\vec{b}_2 = (2, 5)$.
We have $\vec{b}_1\leq \vec{z}_{1, 1}=(3,10)$ and $\vec{b}_2\leq \vec{z}_{2, 1}=(2,5)$; since $\vec{b}_1\geq \vec{b}_2$, we have $d_1 , d_2 = (1,0)$. The other solutions are as shown in Table \ref{tab:solution}. \hfill 
     $\triangle$
    
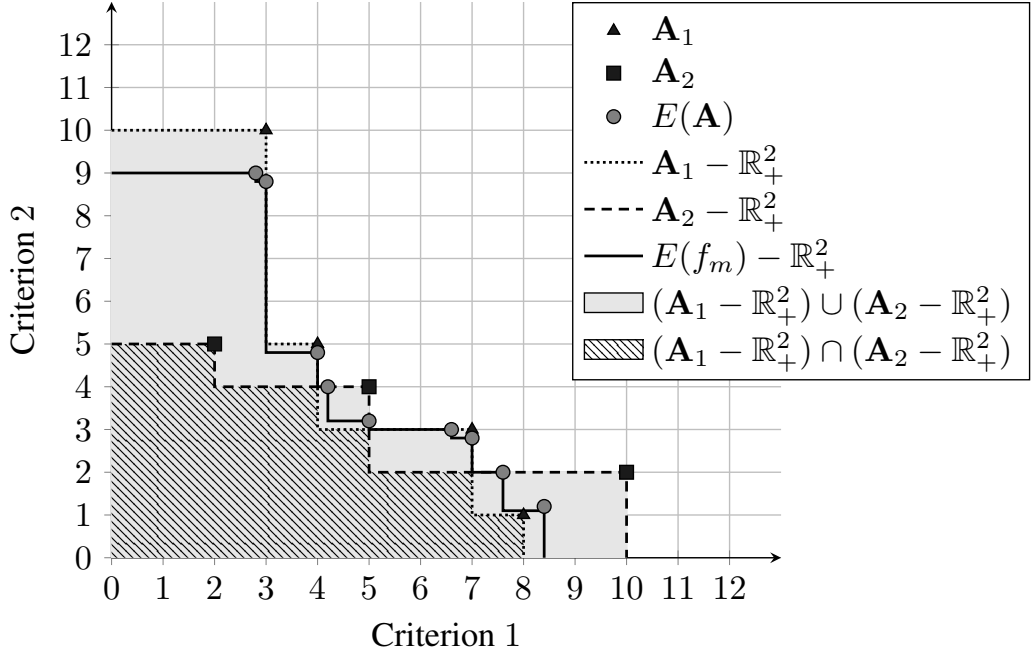
\begin{figure}[htbp]
\centering
\normalsize
\resizebox{\textwidth}{!}{%
\begin{tikzpicture}
\begin{axis}[
    title={},
    xlabel={Criterion $1$},
    ylabel={Criterion $2$},
    xmin=0, xmax=13,
    ymin=0, ymax=13 ,
    xtick={0, 1, 2, 3, 4, 5, 6, 7, 8, 9, 10, 11, 12},
    ytick={0, 1, 2, 3, 4, 5, 6, 7, 8, 9, 10, 11, 12},
    grid=major,
    axis x line=bottom, axis y line = left,
    legend style={at={(1.03,1)},anchor=north},
    legend cell align={left}]

\fill[black!10] (0,0) rectangle (3,10);
\fill[black!10] (3,0) rectangle (4,5);
\fill[black!10] (4,0) rectangle (5,4);
\fill[black!10] (5,0) rectangle (7,3);
\fill[black!10] (7,0) rectangle (10,2);

\draw[draw=none, pattern=north west lines] (0, 0)--(0, 5)--(2, 5)--(2, 4)--(4, 4)--(4, 3)--(5, 3)--(5, 2)--(7, 2)--(7, 1)--(8, 1)--(8, 0)--(0, 0) -- cycle;
\addplot[fill=black!90, mark=triangle*, only marks]
coordinates {(3, 10)(4, 5)(7, 3)(8, 1)};

\addplot[fill=black!90, mark = square*, only marks]
coordinates {(2, 5) (5, 4) (10, 2)};

\addplot[fill=black!50, mark = *, only marks]
coordinates {(2.8, 9) (3, 8.8) (4, 4.8) (4.2, 4) (5, 3.2) (6.6,3)  (7, 2.8) (7.6,2) (8.4, 1.2)};

\addplot[densely dotted, thick
]
coordinates {(0, 10)(3, 10)(3,5)(4,5)(4,3)(7, 3)(7,1)(8,1)(8,0)};

\addplot[densely dashed,thick
]
coordinates {(0, 5)(2, 5)(2,4)(5,4)(5,2)(10, 2)(10,0)};

\addplot[thick
]
coordinates {(0,9)(2.8, 9)(2.8, 8.8) (3, 8.8)(3,4.8) (4, 4.8) (4,4) (4.2, 4) (4.2, 3.2)(5, 3.2)(5, 3) (6.6,3) (6.6,2.8) (7, 2.8)(7, 2)(7.6,2) (7.6,1.1)(8.4, 1.1)(8.4, 0)};
\footnotesize
\addlegendentry{$\A_1$};
\addlegendentry{$\A_2$};
\addlegendentry{$E(\A)$};
\addlegendentry{$\A_1 - \mathbb{R}^2_+$};
\addlegendimage{area legend,fill = black, fill opacity=0.1}
\addlegendimage{area legend,pattern=north west lines}
\addlegendentry{$\A_2 - \mathbb{R}^2_+$};
\addlegendentry{$E(f_m) - \mathbb{R}^2_+$};
\addlegendentry{$(\A_1 - \mathbb{R}^2_+) \cup (\A_2 - \mathbb{R}^2_+)$};
\addlegendentry{$(\A_1 - \mathbb{R}^2_+) \cap (\A_2 - \mathbb{R}^2_+)$};

\end{axis}
\end{tikzpicture}}
    \caption{Example \ref{para:ex2}: The expected capability set $E(\A)$ with $\mathbf{p(s_1) = 0.8}$ and $\mathbf{p(s_2) = 0.2}$}
    \label{fig:example_mix}
\end{figure}
\end{ex}

\begin{table}[htbp]
    \centering
    \begin{tabular}{l||lcc|lcc|c}
      $\vec{b}\in E(\A)  $ & $\vec{z_1} $ & $\;\;$ & $\vec{b_1} $  & $\vec{z_2} $ & $\;\;$ & $\vec{b_2}$ & $(d_1, d_2)$  \\
         $(2.8, 9)$ & $(3, 10)$ && $(3, 10)$ & $(2, 5)$ && $(2, 5)$ &$(1, 0)$\\
         $(3, 8.8)$   & $(3, 10)$ && $(3, 10)$ & $(5, 4)$ && $(3, 4)$ & $(1,0)$\\
         $(4, 4.8)$   & $(4, 5)$ && $(4, 5)$ & $(5, 4)$ && $(4, 4)$ & $(1, 0)$\\
         $(4.2, 4)$   & $(4, 5)$ && $(4, 4)$ & $(5, 4)$ && $(5, 4)$ & $(0, 1)$\\
         $(5, 3.2)$   & $(7, 3)$ && $(5, 3)$ & $(5, 4)$ && $(5, 4)$ & $(0, 1)$\\
         $(6.6,3)$    & $(7, 3)$ && $(7, 3)$ & $(5, 4)$ && $(5, 3)$ & $(1, 0)$\\
         $(7, 2.8)$   & $(7, 3)$ && $(7, 3)$ & $(10, 2)$ && $(7, 2)$ & $(1, 0)$\\
         $(7.6,2)$    & $(7, 3)$ && $(7, 2)$ & $(10, 2)$ && $(10, 2)$ & $(0, 1)$\\
         $(8.4, 1.2)$ & $(8, 1)$ && $(8, 1)$ & $(10, 2)$ && $(10, 2)$ & $(0,1)$\\
    \end{tabular}
    \caption{Example \ref{para:ex2}: Calculation of $\mathbf{E(\A)}$}
    \label{tab:solution}
\end{table}



\vspace{.05in} 

\addtocounter{ex}{-2}
\begin{ex}[Cont]

Figure \ref{fig:E4} depicts the expected capability set $E(f_m)$ when $p(s_1)=p(s_2)=0.5$. Notably, $E(\A)$ is situated between the set of solutions dominated across all states, i.e., $(\A_1 - \mathbb{R}^2_+) \cap (\A_2 - \mathbb{R}^2_+) \subseteq E(\A)$, and the set of solutions dominated by at least one state of the world, i.e., $E(\A)\subseteq (\A_1 \cup \A_2) - \mathbb{R}^2_+$.  \hfill $\triangle$

\begin{figure}[htbp]
\centering
\begin{tikzpicture}
\begin{axis}[
    title={},
    xlabel={Criterion 1},
    ylabel={Criterion 2},
    xmin=0, xmax=10,
    ymin=0, ymax=10 ,
    xtick={0, 1, 2, 3, 4, 5, 6, 7, 8, 9},
    ytick={0, 1, 2, 3, 4, 5, 6, 7, 8, 9},
    grid=major,
    axis x line=bottom, axis y line = left,
    legend style={at={(1.03,1)},anchor=north},
    legend cell align={left}]
 \draw  [gray!20,fill=gray!20] (0.01,0.01) -- (0.01,2) -- (7,2)--(7,0.01) -- cycle;
  \draw [gray!20,fill=gray!20] (0.01,0.01) -- (0.01,3) -- (4,3)--(4,0.01) -- cycle;
\draw [gray!20,fill=gray!20] (0.01,0.01) -- (0.01,3) -- (4,3)--(4,0.01) -- cycle;
 \draw [gray!20,fill=gray!20] (0.01,0.01) -- (0.01,7) -- (2,7)--(2,0.01) -- cycle;
  \draw [gray!20,fill=gray!20] (0.01,0.01) -- (0.01,4) -- (3,4)--(3,0.01) -- cycle;
    \draw[gray!20, fill=gray!50] (0,0) rectangle (3,3);
  \draw[gray!20, pattern=dots] (0,0) rectangle (2,5);
  \draw[gray!20, pattern=dots] (0,0) rectangle (3,3.5);
  \draw[gray!20, pattern=dots] (0,0) rectangle (3.5,3);
  \draw[gray!20, pattern=dots] (0,0) rectangle (5,2);
\draw[gray!50] (2,0)--(2,3);
\draw[gray!50] (0, 2)--(3,2);
\addplot[fill=black!90, mark = triangle*, only marks] coordinates {(2, 7)(3, 4)};
\addplot[fill=black!90, mark = square*, only marks] coordinates {(7,2)(4,3)};
\addplot[fill=black!50, mark = *, only marks] coordinates {(2,5)(3,3.5)(3.5,3)(5,2)};

\draw (8.5,-4) node {$u_1$};
\draw (-4,8.5) node {$u_2$};
\footnotesize
\addlegendentry{$\A_1$};
\addlegendentry{$\A_2$};
\addlegendentry{$E(\A)$};
\addlegendimage{area legend,fill=gray!20}
\addlegendimage{area legend,fill=gray!50}
\addlegendimage{area legend,pattern=dots}
\addlegendentry{$(\A_1 - \mathbb{R}^2_+) \cup (\A_2 - \mathbb{R}^2_+)$};
\addlegendentry{$(\A_1 - \mathbb{R}^2_+) \cap (\A_2 - \mathbb{R}^2_+)$};
\addlegendentry{$E(\A) - \mathbb{R}^2_+$};
\end{axis}

\end{tikzpicture}
    \caption{Example \ref{ex:1}: The expected capability set $\mathbf{E(\A)}$}
    \label{fig:E4}
\end{figure}
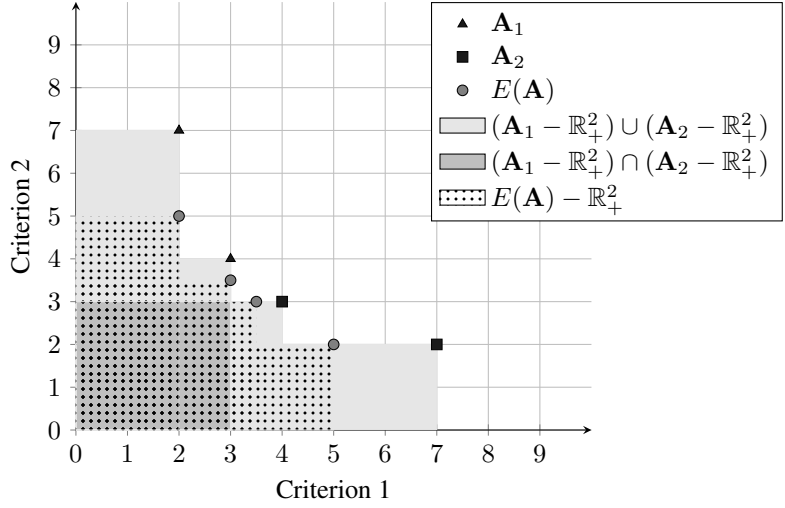
\end{ex}
\addtocounter{ex}{+2}

\section{Properties of expected capability sets}\label{properties}

Inspired by \cite{savage1972foundations} axioms, we
 consider the major properties satisfied by the proposed expected capability concept. 
For comparison, properties for average capability sets are provided in the appendix.

  The first property expresses consistency with the classical maximum expected utility principle under the original conditions 
 in Savage's setup (univariate utilities and single consequences). 
 \vspace{.05in}
 
\begin{prop}[Consistency with expected utility]\label{prop:mix_Consistency}
If the capability sets include just one beings and are only assessed using one dimension, then expected capability sets are equivalent to expected utilities.
\end{prop}
\noindent\textbf{Proof}
Suppose capability sets 
 include only one beings ($|\A_l|= 1$ for all $s_l$) and there is just 
 one assessment dimension ($h^*=1$). In such case, we can express $\A_l$ as a single real number, $b_l$.
Given the  total order among all $b_l$, and considering that there is 
a unique solution for each state, it follows that $b = \max \sum^{l^*}_{l=1} p(s_l) \cdot (b_l - \mathbb{R}_+)$ if and only if $b=\sum^{l^*}_{l=1} p(s_l) \cdot b_l$.
This implies that $b$ is the expected utility of $f_m$, denoted $\sum^{l^*}_{l=1} p(s_l) \cdot \A_l$.
\hfill$\blacksquare$\\

The second property, derived from Property \ref{ax:union}, states that the expected capability set should be weakly Pareto dominated by the union of capability sets across all states. This property is deemed desirable because if PMs are certain that they cannot dominate solution $\vec{b}$ (i.e., for all states, there is no $\vec{b^\prime}\in \A_l$ such that $\vec{b} \leq \vec{b^\prime}$), they should not expect to dominate $\vec{b}$.
By adhering to this property, PMs maintain a realistic understanding of the feasible choices and avoid including beings that citizens are certain to not dominate regardless of the  state of nature. 
To illustrate this, consider the case of a one-dimensional expected capability set, where each capability set $\A_l$ is one-dimensional and contains only one beings (i.e., $\A_l= \{ b_l \} $ with $b_l\in \mathbb{R}$). 
  Under this setting, the expected capability set is dominated by the best solution among all capability sets from different states. 
In other words, the expected capability set is dominated by the union of all capability sets. 
Mathematically, this is expressed as $E(\A) \leq \max_{\forall s_l} U(f_m(s_l))$ or, equivalently $E(\A) \in \bigcup_{l=1}^{l^*} (\A_l-\mathbb{R}_+)$.
\vspace{.05in}

\begin{prop}[Sure domination of the Expected Capability]\label{prop:mix_no_dom_impo_cons}
The expected capability set is dominated by the union of capability sets over all possible states, that is, $$E(\A) \subseteq \bigcup_{l=1}^{l^*} (\A_l - \mathbb{R}^{h^*}_+)$$
\end{prop}
\noindent\textbf{Proof } Consider any solution $\vec{b}$ in $E(\A)$ obtained by aggregating $\sum_{l=1}^{l^*} p(s_l)\cdot \vec{b_l}$. By definition, there is an order among all the beings $\vec{b_l}$ used in this aggregation. Therefore, $\vec{b}$ is at least weakly Pareto dominated by the best beings used in the aggregation.
This highest beings is necessarily weakly Pareto dominated by a solution in $\A_l$. \hfill$\blacksquare$  \\

The next property derives from property \ref{ax:inter} and specifies that the expected capability set should dominate the solutions dominated in the capability set in every state. This 
 seems desirable since if $\vec{b} \in \bigcap_{l=1}^{l^*} (\A_l - \mathbb{R}^{h^*}_+)$, we can find a beings that dominates $\vec{b}\,\,$ under every state. 
To illustrate this, consider the previous one dimension and one beings capability setting. 
In this state of the world, the expected capability set dominates the worst solution among all capability sets from different states. 
In other words, the expected capability set dominates the intersection of the space dominated by capability sets in every state. Mathematically, this is expressed as $\min_{\forall s_l} U(f_m(s_l))\leq E(\A)$ or equivalently $\bigcap_{l=1}^{l^*} (\A_l-\mathbb{R}_+) \in E(\A) -\mathbb{R}_+$.
\begin{prop}[Sure domination by the Expected Capability]\label{prop:mix_sure}
  Expected capability sets dominate the intersection of all beings dominated by all states' capability sets, that is 
$$ \bigcap_{l=1}^{l^*} (\A_l- \mathbb{R}^{h^*}_+) \subseteq E(\A)-\mathbb{R}^{h^*}_+ $$
\end{prop}
\textbf{Proof}
Consider any beings $\vec{b}$ that is dominated by every capability set, i.e.\ $\vec{b}\in \bigcap_{l=1}^{l^*}(\A_l - \mathbb{R}^{h^*}_+)$. For each $l \in \{1, \ldots, l^*\}$, there exists $ \vec{b}_l \in (\A_l-\mathbb{R}^{h^*}_+)$ such that $\vec{b}_l = \vec{b}$.
Therefore, the solution $\sum_{i=l}^{l^*} p(s_l)\cdot \vec{b}_l\in E(\A)$ is equal to $\vec{b}$. \hfill $\blacksquare$\\

The fourth one shows that the expected procedures are preserved by positive affine transformations. This relates to the positive affine uniqueness property of utility functions \citep{french}.
\vspace{.05in}

\begin{prop}[Linearity]\label{prop:mix_linea}
$(a)$ Preservation of addition:
$$E(\A + \vec{c}\;) =E(\A) + \vec{c} \text{ with } \vec{c}\in \mathbb{R}^{h^*}$$
$(b)$ Preservation of positive multiplication 
$$E(\A\cdot \vec{c}\;) = E(\A)\cdot \vec{c} \text{ with } \vec{c}\in \mathbb{R}^{h^*}_+$$ 
\end{prop}
\noindent\textbf{Proof}
To prove $(a)$, consider any $\vec{b}$ in $E(\A + \vec{c}\;)$. This means $\vec{b} = \sum_{l=1}^{l^*} (p(s_l) \cdot \vec{b}_l)$, with an existing order between all $\vec{b}_l$, and $\vec{b}_l \in \A_l + \vec{c} - \mathbb{R}^{h^*}_+$. Similarly, for any $\vec{b}^{\prime}$ in $E(\A) + \vec{c}$, we have $\vec{b}^{\prime} \in \sum_{l=1}^{l^*} (p(s_l) \cdot \vec{b}^\prime_{l}) + \vec{c}$, with an order between all $\vec{b}^\prime_{l}$, and $\vec{b}^\prime_{l} \in \A_l - \mathbb{R}^{h^*}_+$.
Now, for each $\vec{b}$ in $E(\A + \vec{c}\;)$, we can find a $\vec{b}^{\prime}$ in $E(\A)+\vec{c}$ such that $\vec{b}^{\prime}=\vec{b}$, by setting $\vec{b}^\prime_{l} = \vec{b}_l - \vec{c}$.
Conversely, for each $\vec{b}^{\prime}$ in $E(f_m) + \vec{c}$, we can find a $\vec{b} \in E(\A + \vec{c}\;)$ such that $\vec{b}=\vec{b}^{\prime}$, by setting $\vec{b}_l = \vec{b}^\prime_{l} + \vec{c}$.

The proof for $(b)$ follows a similar approach.
\hfill $\blacksquare$ \\

The fifth property stipulates that extending capability sets on states cannot lead to a reduction in the expected capability set. 
\vspace{.05in}

\begin{prop}[Monotonicity over capability domination]\label{prop:mix_mono}
If for all $s_l \in S$ we have $\A_l \subseteq [\B_l-\mathbb{R}^{h^*}_+]$,  then $E(\A) \subseteq E(\B) - \mathbb{R}^{h^*}_+$. 
\end{prop}

\noindent\textbf{Proof }
 For every beings $\vec{b}$ in $E(\A)$, where $\vec{b} = \sum_{l=1}^{l^*} p(s_l)\cdot \vec{b}_l$, there exists a $\vec{b}^{\prime}= \sum_{l=1}^{l^*} p(s_l)\cdot \vec{b}_l^{\prime}$ in $E(\B)-\mathbb{R}^{h^*}_+$ such that $\vec{b}^{\prime} =\vec{b}$ by setting $\vec{b}_l = \vec{b}_l^{\prime}$ for all $l\in \{1, \ldots, l^*\}$.  \hfill $\blacksquare$ \\

Our sixth property shows that increasing the probability of one state of the world dominating another 
one should lead to the expected capability set of the first one to dominate the expected capability set of the second one.
\vspace{.05in}

\begin{prop}[Monotonicity over probabilities]\label{prop:mix_monoprob}
If we have $\A_z \subseteq \A_{z^\prime}-\mathbb{R}^{h^*}_+$ with $(s_z, s_{z'})\in S^2$ and \\
 $p^\prime(s_l) = 
\begin{cases}
p(s_l) + c \;\;\;\; \text{, for}\;\;\;\;l= z'\\
p(s_l) - c \;\;\;\;\text{, for}\;\;\;\; l= z\\
p(s_l)\;\;\;\;\;\;\;\;\;\; \text{, otherwise}
\end{cases}$ \\
with $c \in (0;p(s_z)]$.  
Then, 
$E(\A, p) \subseteq E(\A, p^\prime)) -\mathbb{R}^{h^*}_+$.
\end{prop}
\noindent\textbf{Proof}
 For every beings $\vec{b}$ in $E(\A, p)$, defined as $\vec{b} = \sum^{l^*}_{l=1} p(s_l) \cdot \vec{b}_l$, there always exists a corresponding beings $\vec{b}^{\prime}$ in $E(\A, p^{\prime})$, defined as $\vec{b}^\prime = \sum^{l^*}_{l=1} p^\prime(s_l) \cdot \vec{b^\prime_l}$, such that $\vec{b}^{\prime} \geq \vec{b}$:
\begin{itemize}

    \item If $\vec{b}_z \leq \vec{b}_{z'}$, let $\vec{b}_l = \vec{b}_l^\prime$ for all $l\in \{1, \ldots, l^*\}$. It then follows that $\sum^{l^*}_{l=1} p^\prime(s_l) \cdot \vec{b}_l^\prime \geq \sum^{l^*}_{l=1} p(s_l) \cdot \vec{b}_l$.

    \item If $\vec{b}_z > \vec{b}_{z'}$, let $\vec{b}_l^\prime$ be defined as:

$$\vec{b}_l^\prime = \begin{cases}
\vec{b}_z \;\;\;\; \text{for}\;\; l = z' \\
\vec{b}_{z'} \;\;\;\; \text{for}\;\; l = z 
\\
\vec{b}_l \;\;\;\; \text{Otherwise}\end{cases}$$
 Then, we can deduce that $\sum^{l^*}_{l=1} p^\prime(s_l) \cdot \vec{b}_l^\prime \geq \sum^{l^*}_{l=1} p(s_l) \cdot \vec{b}_l$.\hfill $\blacksquare$ \\

\end{itemize}
 


\section{Expected vs Average Capability sets}\label{seq:vs}

Average capability sets respect the same properties as expected capability sets except for the no domination of impossible beings (Proposition \ref{prop:mix_no_dom_impo_cons}), and the monotonicity over probabilities (Proposition \ref{prop:mix_monoprob}), see the Appendix for proofs.

An interesting relation between both concepts is that for all solutions in the expected capability set, 
it is possible to find a solution in the average capability set that is at least as preferred as it, i.e.\ the expected capability set is dominated by the average capability set. 

\vspace{.05in} 

\begin{prop}
We have
\[ E(\A)\subseteq \mathbf{\overline{\A}} -\mathbb{R}^{h^*}_+ \]
\end{prop}
\noindent \textbf{Proof.}  
For every beings $\vec{b}$ in $E(\A)$ such that $\vec{b}= \sum^{l^*}_{l=1} p(s_l)\cdot \vec{b}_l$, we have that for each $\vec{b}_l$ there exists a corresponding beings $\vec{b}^\prime_l$ in $\A_l$ such that $\vec{b}^\prime_l\geq \vec{b}_l$. Consequently, the beings $\vec{b}^\prime = \sum^{l^*}_{l=1} p(s_l)\cdot \vec{b}^\prime_l$ is in $\overline{\A}$ and satisfies $\vec{b}^\prime \geq \vec{b}$.\hfill $\blacksquare$\\

\begin{figure}
    \centering
    \footnotesize
    \setlength{\tabcolsep}{0pt} 
    \begin{tabular}{m{.33\textwidth}m{.33\textwidth}m{.33\textwidth}}
        \multicolumn{3}{c}{SET 1}\\
        \begin{minipage}{.33\textwidth}
          \includegraphics[width=.98\linewidth]{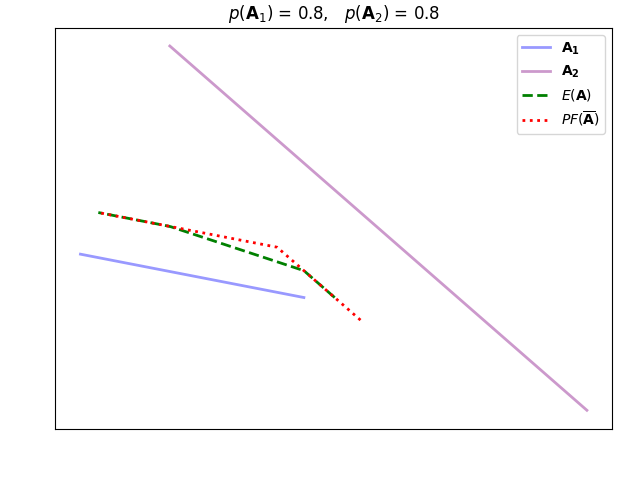}
        \end{minipage} &
        \begin{minipage}{.33\textwidth}
          \includegraphics[width=.98\linewidth]{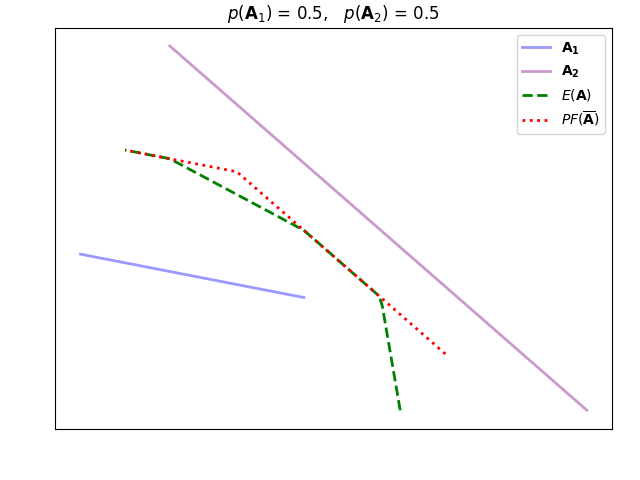}
        \end{minipage} &
        \begin{minipage}{.33\textwidth}
          \includegraphics[width=.98\linewidth]{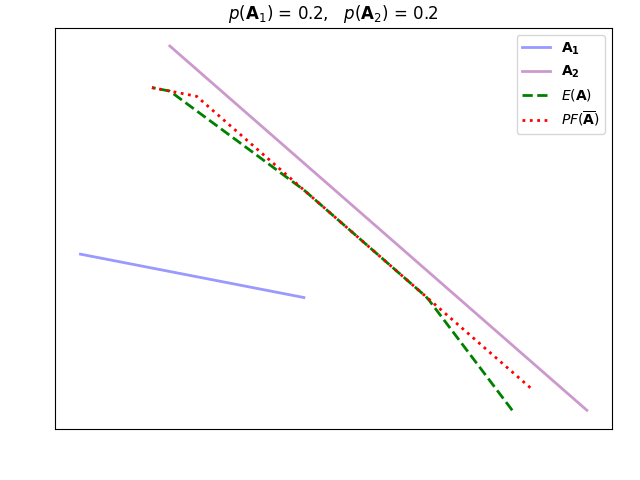}
        \end{minipage}\\\hline

        \multicolumn{3}{c}{SET 2}\\
        \begin{minipage}{.33\textwidth}
          \includegraphics[width=.98\linewidth]{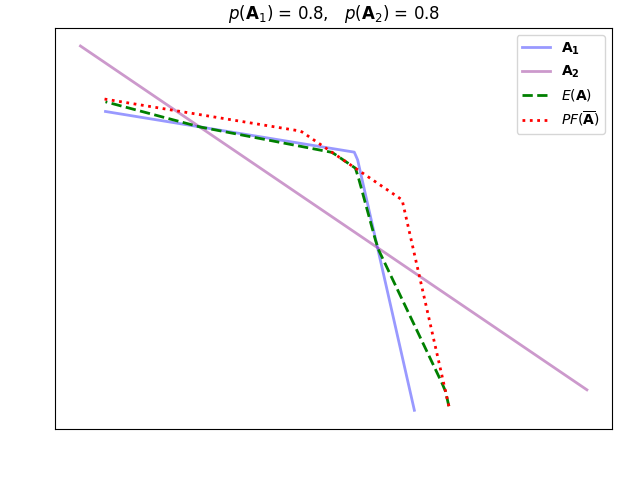}
        \end{minipage} &
        \begin{minipage}{.33\textwidth}
          \includegraphics[width=.98\linewidth]{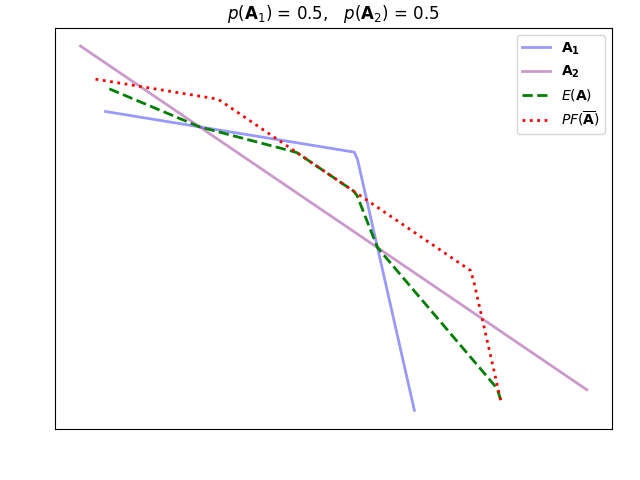}
        \end{minipage} &
        \begin{minipage}{.33\textwidth}
          \includegraphics[width=.98\linewidth]{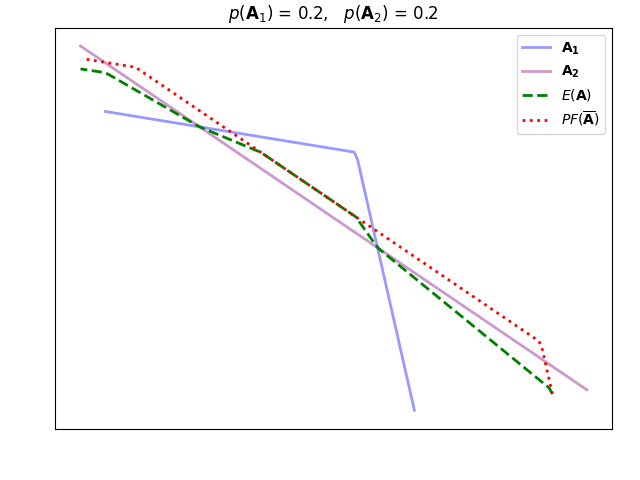}
        \end{minipage}\\\hline

        \multicolumn{3}{c}{SET 3}\\
        \begin{minipage}{.33\textwidth}
          \includegraphics[width=.98\linewidth]{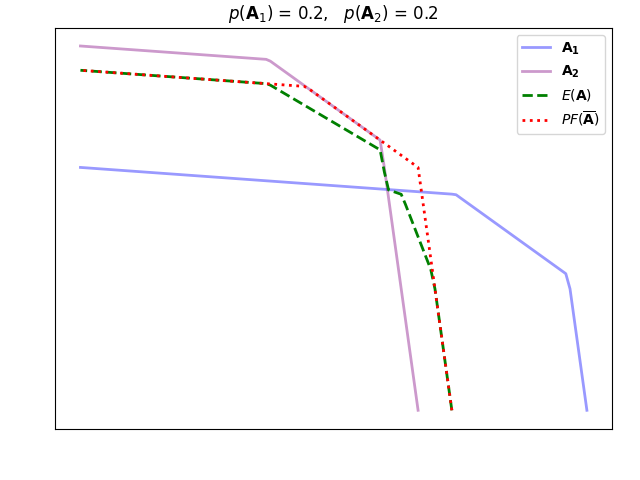}
        \end{minipage} &
        \begin{minipage}{.33\textwidth}
          \includegraphics[width=.98\linewidth]{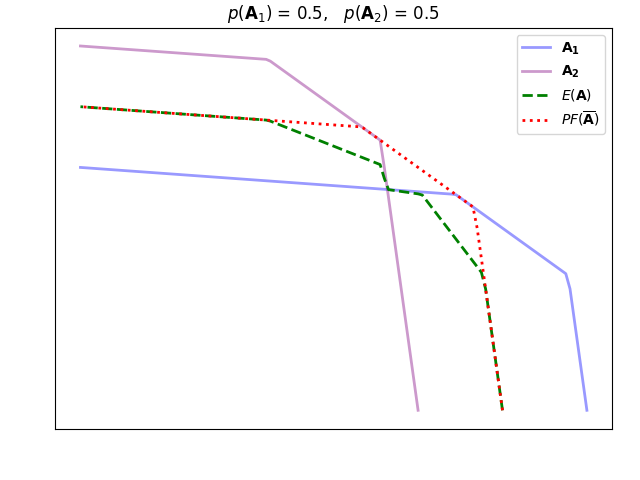}
        \end{minipage} &
        \begin{minipage}{.33\textwidth}
          \includegraphics[width=.98\linewidth]{f32.png}
        \end{minipage}\\
    \end{tabular}
    \caption{Some examples of expected and average capability sets}
    \label{fig:myLboro}
\end{figure}

To illustrate the distinction both sets,  consider the example of mixing two capability sets as Figure \ref{fig:myLboro} shows. In both frameworks, the mixed capability set tends to align more closely with $\A_1$ as the probability $p(s_1)$ increases, and similarly with $\A_2$ as its probability $p(s_2)$ increases.
Furthermore, observe that the expected capability sets are always ``below'' the union of $\A_1$ and $\A_2$, as well as ``below'' the average capability set $PF(\overline{\A})$.

Expected and average capability sets therefor fulfill different properties. Thus, a careful scrutiny is necessary to determine which set to employ under given circumstances.
The expected capability approach is best suited when calculating the anticipated expected beings of a citizen. In this context, Proposition \ref{prop:mix_no_dom_impo_cons} becomes 
 highly relevant. Indeed, it is crucial that no citizen should expect to dominate beings that are not weakly Pareto dominated in any state of the world. 
 Moreover, when increasing the probabilities of a state of the world leading to a capability set that is preferred over another one generated in a different state (Proposition \ref{prop:mix_monoprob}), it is crucial that the resulting expected capability set is improved for the citizen.
 
Conversely, the average capability approach is ideal in situations requiring social aggregation. In this approach, the average capability set of a group reflects a compound of all possible beings that any citizen within the group could choose. Therefore, it is not surprising that average capability sets do not fulfill Proposition \ref{prop:mix_no_dom_impo_cons}. This is because the ``social beings'' in an average capability set represent a combination of ``individual beings''.
This is also why we recommend compiling \( \overline{\A} \) and not necessarily \( PF(\overline{\A}) \) because \( \overline{\A} \) embodies actual social aggregation. This aggregation may not be efficient in the sense that individuals maximize their beings regardless of the aggregated beings at the social level, while \( PF(\overline{\A}) \) would represent the social aggregation where all individuals choose in accordance with an efficient solution at the social level.
The non-fulfillment of Proposition \ref{prop:mix_monoprob} is expected because increasing the probability of a better capability set will make the average capability set ``closer'' to it, but not necessarily dominate the previous average capability set.

We illustrate such distinctions with two final examples.
\vspace{.05in}

\addtocounter{ex}{-1}
\begin{ex}\label{ex:box}
Consider a local farmers' market where two farmers set up identical stalls. Each stall offers two baskets for sale. One basket contains 2 apples and 4 carrots; the other holds 4 apples and 2 carrots.
Two shoppers, Alice and Bob, visit the market with enough money to purchase only one basket each. The act $f_m$ represents the random assignment of Alice and Bob to one of these stalls (we obtain $\A_{alice} = \A_{bob} = \{(apples:2, carrots:4), (apples:4, carrots:2)\}$). Let us determine the average capability set in purchasing apples and carrots in the community ($\overline{\A}=\{(apples: 2, carrots: 4),(apples: 3, carrots: 3),(apples: 4, carrots: 2)\}$). Then,\\
- If both Alice and Bob choose the first basket, the average basket of Alice and Bob contains 2 apples and 4 carrots per person.\\
- If Alice and Bob each select different baskets, the average basket of Alice and Bob contains 3 apples and 3 carrots per person.\\
- If both Alice and Bob opt for the second basket, the average basket of Alice and Bob contains 4 apples and 2 carrots per person.

Now, consider the particular case of a shopper that is randomly directed to one of the stalls and is allowed to select one of the baskets (we obtain $\A_1=\A_2=\{(apples: 2, carrots: 4),(apples: 4, carrots: 2)\}$). In this context, the expected capability set for the shopper would be $E(\A)= \{(apples: 2, carrots: 4),(apples: 4, carrots: 2)\}$:\\
- 2 apples and 4 carrots if they select the first basket\\
- 4 apples and 2 carrots if they select the second basket.
$\hfill \triangle$
\end{ex}


\vspace{.05in}

\begin{ex}\label{ex:magic}
Consider a region where the local government has two types of land grants available: Land Grant 1 offers plots in a rain-abundant valley, ideal for rice cultivation. On the other hand, Land Grant 2 provides plots on sun-soaked hill terraces, perfect for vineyards. The government, in an attempt to diversify agriculture, randomly assigns Land Grant 1 to farmer Alice and Land Grant 2 to farmer Bob.

In this scenario, the term ``average capability set'' refers to the expected agricultural products both farmers can yield. Although Alice has the exclusive capability to cultivate rice due to her land's conditions, and Bob has the unique advantage to grow grapes ($\A_{Alice}=\{(rice: [0;M], grapes: 0)\}, \A_{Bob}=\{(rice: 0, grapes: [0;M])\}$ with $M$ a large number), each farmer in this community, on average, possesses the ability to produce both rice and grapes. Therefore, the average capability set align with any reasonable amount of rice and grapes per farmer $\overline{\A}= \{(rice: [0;\frac{M}{2}], grapes: [0;\frac{M}{2}])\}$.

 In contrast, Alice's expected yield is contingent on the specific land grant she receives $\A_{valley}=\{(rice:[0;M], wine:0), \A_{terrace}=\{(rice:0, wine:[0;M])\}$. If she is assigned Land Grant 1, she can cultivate an ample amount of rice but no grapes. However, if she were to receive Land Grant 2, she could produce grapes but not rice. Hence, Alice's expected yield encompasses up to a significant amount of either rice or grapes, but not both, $E(\A) = \{(rice:\frac{M}{2}, wine:0), (rice:0, wine:\frac{M}{2})\}$.$\hfill \triangle$
\end{ex}

\vspace{.05in}

\noindent These examples highlight the critical distinction between average and expected capability sets. While average capability sets reflect the combined possible outcomes in a population, expected capability sets provide a probabilistic forecast of the individual's potential outcomes. The choice between both measures would be context-dependent, whether it requires a societal or an individual perspective.
\section{Conclusion}
\addcontentsline{toc}{section}{Conclusion}
We have provided two mixing concepts that can be used to summarise capability sets in a multiobjective decision-making setup under risk within the capability approach: expected and average capability sets. 
Both have similar properties but are crucially different concerning 
   sure domination of the expected capability (Proposition \ref{prop:mix_no_dom_impo_cons}).
The choice of which concept to apply depends on the decision-maker's ultimate goal. 
Specifically, if the PM is dealing with capability sets linked to states of the world, each having definite probabilities, and only one of these states of the world will materialize, then using expected capability sets is recommended. Conversely, if the situation involves the combination of multiple capability sets, and all beings can be achieved by different individuals, average capability sets seem a more suitable choice.

This general framework of representing sets of choices in a multidimensional space extends beyond the capability approach within welfare economics. 
It can also find relevance and application in other fields and contexts where decision-making involves multiple choice and criteria:

\begin{itemize}

\item In an uncertain environment, capability sets could refer 
to the potential results associated with risk treatments 
before we are able to undertake a full probabilistic assessment of  impacts
 \citep{bossert2000choice,cohen1980rational}.

\item In research and development (R\&D) management, capability sets can refer to the potential outcomes through projects. 
R\&D decisions are often made over multiple periods, where choices made today impact options available in the future. 
The question then arises as to which development option should be chosen today, knowing that it will influence future development choices.
Furthermore, in many cases, the decision maker may not have precise knowledge of their future possible choices due to uncertainties related to factors such as the economy, societal values, environmental conditions, future laws, etc. 
The sets of possible future choices would also be uncertain due to the inherent uncertainties associated with R\&D projects, such as performance, development time, and costs. 
Representing these choices as capability sets would enable flexibility and resilience \citep{evans1991strategic,koopmans1962flexibility,kreps1979representation}, allowing for better adaptation to unforeseen events.

    \item Lastly, in a multi-objective setting \citep{bouyssou2006evaluation}, capability sets  
    would refer to the multiple objective levels achievable
    under various states, when we have not yet undertaken
    a process to aggregate results.
\end{itemize}

From a framework perspective, the proposed setting extends naturally to two important cases.
Firstly, we assumed a multi-objective setting essentially entailing using the whole set of monotonic utility functions (rather than a single utility function as in Savage's framework).
In between, we could conceive cases in which a smaller class of utility functions is used; as an example, we could consider the set of 
all monotonic risk-averse utility functions, as in second order 
stochastic dominance.
Secondly, we assumed a single probability distribution over the states, but again, there could be cases in which we have a class of probability distributions as
in robust Bayesian settings 
\citep{ruggeri}.

Finally, another exciting direction for future research lies in developing new methodologies to assess expected capability sets.
This exploration should go beyond our proposed approach, exploring more nuanced strategies as in \cite{pattanaik1990ranking, foster2011freedom,barbera2004ranking,gaertner2012evaluating,gaertner2006capability,gaertner2008new}. 
The prevailing challenge, which remains unaddressed, lies in finding a way to account for the diversity of beings and how these beings are valued by citizens. 
Advancements in this direction will undoubtedly expand the potential use of capability sets within a decision-making framework, offering further enrichment to this field of study.

\subsection*{Acknowledgement}
Research supported by the AXA-ICMAT Chair in Adversarial RIsk Analysis 
 and the Spanish Ministry of Science program PID2021-124662OB-I00, the Severo Ochoa Excellence Programme CEX-2019-000904-S
and a grant from the FBBVA (Amalfi).



\bibliography{cahier-template}
\begin{appendices}
\section*{Appendix}\label{Annex:Proof}
The appendix provides proofs of the properties satisfied by average capability sets.\\
\vspace{.05in}

\textbf{Proposition \ref{prop:mix_Consistency}bis: Consistency with expected utility}
\emph{
If the capability sets include just one beings and are only assessed using one dimension, then expected capability sets are equivalent to expected utilities.}

\noindent\textbf{Proof}
Suppose we are dealing with 
only one beings ($|\A|= 1$) and one assessment dimension ($h^*=1$). In such a case, we can express $u(f_m(s_l))$ as a single real number, $b_l$.
Then $\overline{\A} =\Bigl\{\sum_{l=1}^{l^*} p(s_l) \cdot X_{i, l}, \text{ for all } X_i $ such that $ X_i = (\vec{b_1}, \ldots , \vec{b_{l^*}}) | \vec{b}_l \in \A_l \text{ for every } l\in \{1, \ldots , l^*\}\Bigr\}$ is equivalent to $\sum_{l=1}^{l^*} p(s_l) \cdot b_l$ which is equivalent to $\sum^{l^*}_{l=1} p(s_l) \cdot u(f_m(s_l))$.
\hfill$\blacksquare$\\
\vspace{.05in}

\noindent
\textbf{Proposition \ref{prop:mix_sure}bis: Sure domination by the average capability}
\emph{The average capability sets include the intersection of all beings dominated by states' capability, that is 
\[ \bigcap_{l=1}^{l^*} (\A_l - \mathbb{R}^{h^*}_+) \subseteq \overline{\A}-\mathbb{R}^{h^*}_+ \] }

\noindent\textbf{Proof }
Consider any beings $\vec{b}$ that is dominated by every capability set, i.e.\ $\vec{b}\in \bigcap_{l=1}^{l^*}(\A_l - \mathbb{R}^{h^*}_+)$. 
For each $l \in \{1, \ldots, l^*\}$, there exists $ \vec{b}_l^\prime \in \A_l$ such that $\vec{b}^\prime_l \geq \vec{b}$.
Therefore, the solution $\sum_{i=l}^{l^*} p(s_l)\cdot X_{i,l}$ with $X_{i}=(\vec{b_1^\prime}, \ldots, \vec{b_{l^*}^\prime})$ is in $\overline{\A}$ and is at least weakly dominating $\vec{b}$. \hfill $\blacksquare$\\
\vspace{.05in}

\noindent\textbf{Proposition \ref{prop:mix_linea}bis: Linearity}\\\emph{
$(a)$ Preservation of addition:
$$\forall \vec{c} \in \mathbb{R}^{h^*},\; \overline{\A+\vec{c}}=\overline{\A}+\vec{c}$$
$(b)$ Preservation of positive multiplication 
$$\forall \vec{c} \in \mathbb{R}^{h^*}_+, \; \overline{\A \cdot \vec{c}}=\overline{\A}\cdot \vec{c} $$}

\noindent\textbf{Proof }To prove $(a)$, we simply remark that\\
\(\begin{aligned}
\overline{\A} + \vec{c} = \biggl\{ &\sum_{l=1}^{l^*} \bigl( p(s_l) \cdot X_{i,l} \bigr) + \vec{c}, 
\text{for all } X_i \text{ such that } X_i = (\vec{b_1}, \ldots, \vec{b_{l^*}})  \\
&\mid \vec{b}_l \in \A_l \text{ for every } l \in \{1, \ldots, l^*\} \biggr\}
\end{aligned}\)\\
and \\
\(\begin{aligned}
\overline{\A + \vec{c}} = \biggl\{ &\sum_{l=1}^{l^*} p(s_l) \cdot X_{i, l},
\text{for all } X_i \text{ such that } X_i = (\vec{b_1}, \ldots, \vec{b_{l^*}})  \\
&\mid \vec{b}_l \in \A_l + \vec{c} \text{ for every } l \in \{1, \ldots, l^*\} \biggr\}
\end{aligned}\)\\
The proof for $(b)$ follows a similar approach.\hfill $\blacksquare$ \\
\vspace{.05in}

\noindent\textbf{Proposition \ref{prop:mix_mono}bis: Monotonicity over capability domination}
\emph{If for all $s_l \in S$ we have $\A_l \subseteq [\B_l-\mathbb{R}^{h^*}_+]$,  then $\overline{\A} \subseteq \overline{\B} - \mathbb{R}^{h^*}_+$. }

\noindent\textbf{Proof } 
 For every beings $\vec{b}$ in $\overline{\A}$, where $\vec{b} = \sum_{l=1}^{l^*} p(s_l)\cdot X_{i,l}$ and $X_{i}=(\vec{b_1}, \ldots, \vec{b_{l^*}})$, there is a set of beings $X^\prime_{i}=(\vec{b_1^\prime}, \ldots, \vec{b_{l^*}^\prime})$ such that $b_l\leq b^\prime_l$ for all $l$ (since for all $s_l \in S$ we have $\A_l \subseteq [\B_l-\mathbb{R}^{h^*}_+]$). 
 Thus $\vec{b}$ is an element of $\vec{b^\prime} - \mathbb{R}^{h^*}_+$ with $\vec{b^\prime} = \sum_{l=1}^{l^*} p(s_l)\cdot X^\prime_{i,l}$ and $\vec{b^\prime}$ is in $\overline{\A^\prime}$.\hfill $\blacksquare$ \\

\vspace{.05in}
We provide now counterexamples for the propositions not fulfilled by average capability sets.

\noindent\textbf{Proposition \ref{prop:mix_no_dom_impo_cons}bis: Possible domination of the average capability} 
\emph{The average capability set is not always dominated by the set of all possible states, that is, we can have $\overline{\A} \nsubseteq \bigcup_{l=1}^{l^*} (\A_l - \mathbb{R}^{l^*}_+)$}.\\
\noindent\textbf{Proof } See Examples \ref{ex:box}, \ref{ex:magic}, or Set 1 and 3 of Figure \ref{fig:myLboro}. \hfill $\blacksquare$\\

\noindent\textbf{Proposition \ref{prop:mix_monoprob}bis: Non-monotonicity over probabilities}\emph{
If we have $\A_z \subseteq \A_{z^\prime}-\mathbb{R}^{h^*}_+$ with $(s_z, s_{z'})\in S^2$ and \\
 $p^\prime(s_l) = 
\begin{cases}
p(s_l) + c \;\;\;\; \text{for}\;\;\;\;l= z'\\
p(s_l) - c \;\;\;\;\text{for}\;\;\;\; l= z\\
p(s_l)\;\;\;\; \text{Otherwise}
\end{cases}$ \\
with $c \in (0;p(s_z)]$.  
Then, we do not necessarily have 
$\overline{\A} \subseteq \overline{\A^\prime} -\mathbb{R}^{h^*}_+$.}

\noindent\textbf{Proof}
Consider a decision problem with two states of the world, $s_1$ and $s_2$, and their associated capability set $\A_1=U(f_m(s_1)) = \{(0, 1)\}$ and $\A_2=U(f_m(s_2)) = \{(0, 1), (1, 0)\}$. We have $\A_1\subseteq \A_2-\mathbb{R}^2_+$

Initially, the policy makers base their decision on $\overline{\A}$ with equal subjective probabilities ($p(s_1) = p(s_2) = 0.5$), resulting in $\overline{\A} = \{(0, 1), (0.5, 0.5)\}$.
 
Now, with increased importance on the ``best'' set (e.i. $\A_2$), the new choice is $\overline{\A^\prime} = \{(0, 1), (0.75, 0.25)\}$ when using subjective probabilities $p^\prime(s_1) = 0.25$ and $p^\prime(s_2) = 0.75$, resulting in $\overline{\A}\nsubseteq\overline{\A^\prime}$. \hfill $\blacksquare$\\
\vspace{.05in}

%
\end{appendices}

\end{document}